\documentstyle[prl,eqsecnum,aps]{revtex}

\begin{document}
\author{Jian-Qi Shen $^{1}$ \footnote{{\bf Electronic address}: jqshen@coer.zju.edu.cn} and Fei Zhuang$^{2}$}
\address{$^{1}$ Centre for Optical
and Electromagnetic Research, State Key Laboratory of Modern
Optical Instrumentation;\\Zhejiang Institute of Modern Physics and
Department of Physics, \\Zhejiang University,
Hangzhou SpringJade 310027, P.R. China\\
$^{2}$ Department of Physics, Hangzhou Teacher's College, Hangzhou
310012, P.R. China}
\date{\today }
\title{Self-induced charge currents in electromagnetic materials, photon effective rest mass and some related topics\footnote{This paper is concerned with
the self-induced charge currents and some related physically
interesting topics such as photon effective rest mass due to {\it
medium dispersion} and its effects in torsion balance experiments.
It will be submitted nowhere else for the publication, just
uploaded at the e-print archives.}} \maketitle

\begin{abstract}
The contribution of self-induced charge currents of metamaterial
media to {\it photon effective rest mass} is discussed in detail
in the present paper. We concern ourselves with two kinds of
photon effective rest mass, {\it i.e.}, the {\it
frequency-dependent} and {\it frequency-independent} effective
rest mass. Based on these two definitions, we calculate the photon
effective rest mass in the left-handed medium and the 2TDLM media,
the latter of which is described by the so-called {\it two time
derivative Lorentz material} (2TDLM) model. Additionally, we
concentrate primarily on the torque, which is caused by the
interaction between self-induced charge currents in dilute plasma
({\it e.g.}, the secondary cosmic rays) and interstellar magnetic
fields (ambient cosmic magnetic vector potentials), acting on the
torsion balance of the rotating torsion balance experiment.
\end{abstract}
\section{Photon effective rest mass due to self-induced charge current}
In is well known that in some simple electromagnetic media such as
electron plasma, superconducting media and Lorentz dispersive
materials (inside which the time-harmonic electromagnetic wave is
propagating), the self-induced charge current density ${\bf
J}$\cite{Ho} is proportional to the magnetic vector potential
${\bf A}$. Thus the interaction term $\mu _{0}{\bf J}\cdot {\bf
A}$ in the Lagrangian density is therefore transformed into an
effective mass term $-\frac{1}{2}\frac{m_{\rm eff}^{2}c^{2}}{\hbar
^{2}}{\bf A}^{2}$ of electromagnetic fields\cite{Shen}, which is
analogous to the rest mass term in the Londons' electromagnetics
for superconductivity, Ginzberg-Landau superconductivity theory
and Higgs mechanism. Note that in these media, the magnetic
permeability\footnote{Note that here for the magnetic properties
in superconductors, we adopt the current viewpoint rather than the
magnetic-charge viewpoint (where the permeability can be viewed as
$\mu=0$).} $\mu=1$, so it is easy for us to obtain ${\bf J}$ (and
hence the {\it effective rest mass} of photon) by solving the
equation of motion of charged particles acted upon by
electromagnetic waves. However, for some artificial composite
electromagnetic metamaterials, which are described by, {\it e.g.},
the {\it two time derivative Lorentz material} (2TDLM) model (in
which both the electric permittivity and the magnetic permeability
are of frequency dependence and therefore of complicated form) and
some uniaxially (or biaxially) anisotropic media with permittivity
and permeability being tensors, the above formulation may be not
helpful in obtaining the {\it effective rest mass} of
electromagnetic fields. In these cases, however, we have shown
that the following formula\cite{Shen}
\begin{equation}
\frac{m_{\rm eff}^{2}c^{4}}{\hbar ^{2}}=(1-n^{2})\omega ^{2},
                       \label{eq1}
\end{equation}
which arises from the Einstein-de Broglie relation, can be
applicable to the above problem. In the expression (\ref{eq1}),
$n$, $\hbar$ and $c$ denote the optical refractive index, Planck's
constant and speed of light in a vacuum, respectively.

The {\it two time derivative Lorentz material} (2TDLM) model first
suggested by Ziolkowski and
Auzanneau\cite{Ziolkowski3,Ziolkowski4} is a generalization of the
standard Lorentz material model. Ziolkowski has shown that this
type of 2TDLM medium can be designed so that it allows
communication signals to propagate in the medium at speeds
exceeding the speed of light in a vacuum without violating
causality\cite{Ziolkowski}. For these artificial metamaterials,
the model encompasses the permittivity and permeability material
responses experimentally obtained\cite{Smith}. The \^{x}-directed
polarization and \^{y}-directed magnetization fields in such
materials would have the following forms\cite{Ziolkowski}
\begin{eqnarray}
\frac{\partial^{2}}{\partial t^{2}}{\mathcal
P}_{x}+\Gamma\frac{\partial}{\partial t}{\mathcal
P}_{x}+\omega^{2}_{0}{\mathcal
P}_{x}=\epsilon_{0}\left({\omega^{2}_{\rm p}\chi^{\rm
e}_{\alpha}{\mathcal E}_{x}+\omega_{\rm p}\chi^{\rm
e}_{\beta}\frac{\partial}{\partial t}{\mathcal E}_{x}+\chi^{\rm
e}_{\gamma}\frac{\partial^{2}}{\partial
t^{2}}{\mathcal E}_{x}}\right),                           \nonumber \\
\frac{\partial^{2}}{\partial t^{2}}{\mathcal
M}_{y}+\Gamma\frac{\partial}{\partial t}{\mathcal
M}_{y}+\omega^{2}_{0}{\mathcal M}_{y}=\omega^{2}_{\rm p}\chi^{\rm
m}_{\alpha}{\mathcal H}_{y}+\omega_{\rm p}\chi^{\rm
m}_{\beta}\frac{\partial}{\partial t}{\mathcal H}_{y}+\chi^{\rm
m}_{\gamma}\frac{\partial^{2}}{\partial t^{2}}{\mathcal H}_{y},
\label{eq2}
\end{eqnarray}
where $\chi _{\alpha }^{\rm e,m}$, $\chi _{\beta }^{\rm e,m}$ and
$\chi _{\gamma }^{\rm e,m}$ represent, respectively, the coupling
of the electric (magnetic) field and its first and second time
derivatives to the local electric (magnetic) dipole moments.
$\omega _{\rm p}$, $\Gamma ^{\rm e}$, $\Gamma ^{\rm m}$ and
$\omega _{0}$ can be viewed as the plasma frequency, damping
frequency and resonance frequency of the electric (magnetic)
dipole oscillators, respectively. Thus the frequency-domain
electric and magnetic susceptibilities are given
\begin{equation}
\chi ^{\rm e}\left( \omega \right)=\frac{\left( \omega _{\rm
p}^{\rm e}\right) ^{2}\chi _{\alpha }^{\rm e}+i\omega \omega _{\rm
p}^{\rm e}\chi _{\beta }^{\rm e}-\omega ^{2}\chi _{\gamma }^{\rm
e}}{-\omega ^{2}+i\omega \Gamma ^{\rm e}+\left( \omega _{0}^{\rm
e}\right) ^{2}}, \quad   \chi ^{\rm m}\left( \omega
\right)=\frac{\left( \omega _{\rm p}^{\rm m}\right) ^{2}\chi
_{\alpha }^{\rm m}+i\omega \omega _{\rm p}^{\rm m}\chi _{\beta
}^{\rm m}-\omega ^{2}\chi _{\gamma }^{\rm m}}{-\omega ^{2}+i\omega
\Gamma ^{\rm m}+\left( \omega _{0}^{\rm m}\right) ^{2}},
\label{eqq4}
\end{equation}
So, the refractive index squared in the 2TDLM model reads
\begin{equation}
n^{2}\left( \omega \right) =1+\chi ^{\rm e}\left( \omega \right)
+\chi ^{\rm m}\left( \omega \right) +\chi ^{\rm e}\left( \omega
\right) \chi ^{\rm m}\left( \omega \right).
\label{eq3}
\end{equation}
Thus it follows from (\ref{eq1}) that the {\it photon effective
rest mass squared} in the 2TDLM model is of the form
\begin{equation}
m_{\rm eff}^{2}=\frac{\hbar ^{2}\omega^{2}}{c^{4}}\left[{\chi
^{\rm e}\left( \omega \right) +\chi ^{\rm m}\left( \omega \right)
+\chi ^{\rm e}\left( \omega \right) \chi ^{\rm m}\left( \omega
\right)}\right]. \label{eqq42}
\end{equation}
As an illustrative example, we apply Eq.(\ref{eqq42}) to the case
of left-handed media, which is a kind of artificial composite
metamaterials with negative refractive index in the microwave
frequency region ($10^{9}$ Hz). More recently, this medium
captured considerable attention in various fields such as
condensed matter physics, materials science, applied
electromagnetics and
optics\cite{Smith,Veselago,Klimov,Pendry3,Shelby,Ziolkowski2}. The
most characteristic features of left-handed media are: ({\rm i})
both the electric permittivity $\epsilon$ and the magnetic
permeability $\mu$ are negative; ({\rm ii}) the Poynting vector
and wave vector of electromagnetic wave propagating inside it
would be antiparallel, {\it i.e.}, the wave vector {\bf {k}}, the
electric field {\bf {E}} and the magnetic field {\bf {H}} form a
{\it left-handed} system; ({\rm iii}) a number of peculiar
electromagnetic and optical properties such as the reversal of
both the Doppler shift and the Cherenkov radiation, anomalous
refraction, amplification of evanescent waves\cite{Pendryprl},
unusual photon tunneling\cite{Zhang}, modified spontaneous
emission rates and even reversals of radiation pressure to
radiation tension\cite{Veselago,Klimov} arise. All these
dramatically different propagation characteristics stem from the
sign change of the group velocity. In what follows we calculate
the {\it photon effective rest mass squared} by making use of
Eq.(\ref{eqq42}) and the expressions
\begin{equation}
\epsilon \left( \omega \right) =1-\frac{\omega _{\rm
p}^{2}}{\omega \left( \omega +i\gamma \right) },    \quad  \mu
\left( \omega \right) =1-\frac{F\omega ^{2}}{\omega ^{2}-\omega
_{0}^{2}+i\omega \Gamma } \label{eqq}
\end{equation}
for the effective dielectric parameter and magnetic permeability,
where the plasma frequency $\omega _{\rm p}$ and the magnetic
resonance frequency $\omega _{0}$ are in the GHz range ({\it
e.g.}, $\omega _{\rm p}=10.0$ GHz, $\omega _{0}=4.0$ GHz
\cite{Ruppin}); $\gamma$ and $\Gamma$ represent the damping
parameters. The parameter $F$ may often be chosen $0<F<1$, for
example, $F=0.56$\cite{Ruppin}. By using $n^{2}\left( \omega
\right) =\epsilon \left( \omega \right) \mu \left( \omega \right)
$, we obtain\cite{Jianqi}
\begin{equation}
n^{2}\left( \omega \right) =1-\frac{\omega _{\rm p}^{2}}{\omega
^{2}}\left( 1+\frac{i\gamma }{\omega }\right) ^{-1}+\frac{F\omega
_{\rm p}^{2}}{\omega ^{2}}\left( 1+\frac{i\gamma }{\omega }\right)
^{-1}\left( 1-\frac{\omega _{0}^{2}}{\omega ^{2}}+\frac{i\Gamma
}{\omega }\right) ^{-1}-F\left( 1-\frac{\omega _{0}^{2}}{\omega
^{2}}+\frac{i\Gamma }{\omega }\right) ^{-1}. \label{eq57}
\end{equation}
It follows that the mass squared $m_{\rm eff}^{2}$ is obtained
\begin{equation}
m_{\rm eff}^{2}=\frac{\hbar ^{2}}{c^{4}}\left[{\omega _{\rm
p}^{2}\left( 1+\frac{i\gamma }{\omega }\right) ^{-1}-F\omega _{\rm
p}^{2}\left( 1+\frac{i\gamma }{\omega }\right) ^{-1}\left(
1-\frac{\omega _{0}^{2}}{\omega ^{2}}+\frac{i\Gamma }{\omega
}\right) ^{-1}+F\omega ^{2}\left( 1-\frac{\omega _{0}^{2}}{\omega
^{2}}+\frac{i\Gamma }{\omega }\right) ^{-1}}\right ].
\label{eq244}
\end{equation}
Note that for the electron-plasma medium, where the
electromagnetic parameters $\gamma=\Gamma=0$, $F=0$, one can
arrive at the following familiar formula
\begin{equation}
m_{\rm eff}^{2}=\frac{\hbar ^{2}\omega _{\rm p}^{2}}{c^{4}}
\label{eq19}
\end{equation}
with the plasma frequency squared $\omega _{\rm
p}^{2}=\frac{Ne^{2}}{\epsilon _{0}m_{\rm e}}$ (here $e$, $m_{\rm
e}$ and $N$ respectively denote the electron charge, mass and
electron number density in this electron plasma).

It should be emphasized that there is a difference in the
definition of {\it photon effective rest mass} between two types
of experimental schemes of testing photon mass\footnote{These two
types of experimental schemes are as follows: one is based on the
Hamiltonian density of electrodynamics and the other based on
Maxwellian equations.}. In the discussion of wave dispersion and
potential variations of the speed of light with frequency, which
are based on the Maxwellian equations ({\it rather than the
Hamiltonian or Lagrangian density of interacting electromagnetic
system}), the effective rest mass squared of photons is expressed
by Eq.(\ref{eq19}). Historically, this problem has been discussed
by many investigators, {\it e.g.}, Feinberg who considered the
possibility of the variation of the speed of light with frequency
in the sharply defined optical and radio pulses from pulsars and
took into account in more detail the observed variation of arrival
time with frequency for the radio waves attributed to the
interaction with interstellar electrons\cite{Feinberg}. However,
for those experimental schemes of testing photon rest mass, which
are based on the Hamiltonian density ({\it rather than the
Maxwellian equations}), the effective rest mass of photons due to
media dispersion ({\it i.e.}, the interaction of wave with charged
particles in environmental dilute plasma such as the secondary
cosmic rays) is two times that expressed in Eq.(\ref{eq19}). The
torsion balance experiment is just this type of schemes, which
will be considered in Sec.V.

However, it should be noted that in Lorentz dispersive materials,
$m_{\rm eff}^{2}$ in (\ref{eq244}) depends strongly on $\omega$,
and sometimes possesses an imaginary part (due to the damping
parameter and resonance frequency) and the physical meanings of
$m_{\rm eff}^{2}$ is therefore not very apparently seen. In what
follows we will define a physically meaningful effective rest mass
of photons, which is independent of the frequency $\omega$, and
will explain a method by which the frequency-independent effective
rest mass of photons in the 2TDLM media is extracted from
$n^{2}\left( \omega \right) $. Note that the above non-physical
disadvantage of frequency-dependent mass in (\ref{eq244}) will be
avoided in the frequency-independent effective rest mass.
\section{Extracting frequency-independent effective mass from refractive index squared}
In this framework of definition of frequency-independent effective
rest mass, the photon should be regarded as acted upon by a
hypothetical force field, namely, it behaves like a massive de
Broglie particle moving in a force field. It is verified in the
following that, for the case of de Broglie particle in a force
field, one can also consider its ``optical refractive index" $n$.
According to the Einstein- de Broglie relation, the dispersion
relation of the de Broglie particle with rest mass $m_{0}$ in the
potential field $V\left( {\bf x}\right) $ agrees with
\begin{equation}
\left( \omega -\phi \right)
^{2}=k^{2}c^{2}+\frac{m_{0}^{2}c^{4}}{\hbar ^{2}} \label{eqA7}
\end{equation}
with $\phi =\frac{V}{\hbar }$. It follows that
\begin{equation}
\omega ^{2}\left[ 1-\frac{m_{0}^{2}c^{4}}{\hbar ^{2}\omega
^{2}}-2\frac{\phi }{\omega }+\left( \frac{\phi }{\omega }\right)
^{2}\right] =k^{2}c^{2},     \label{eqA8}
\end{equation}
which yields
\begin{equation}
\frac{\omega
^{2}}{k^{2}}=\frac{c^{2}}{1-\frac{m_{0}^{2}c^{4}}{\hbar ^{2}\omega
^{2}}-2\frac{\phi }{\omega }+\left( \frac{\phi }{\omega }\right)
^{2}}.                                        \label{eqA9}
\end{equation}
Compared with the dispersion relation $\frac{\omega
^{2}}{k^{2}}=\frac{c^{2}}{n^{2}}$, one can arrive at
\begin{equation}
n^{2}=1-\frac{m_{0}^{2}c^{4}}{\hbar ^{2}\omega ^{2}}-2\frac{\phi
}{\omega }+\left( \frac{\phi }{\omega }\right) ^{2},
                                   \label{eq34}
\end{equation}
which is the square of ``optical refractive index" of the de
Broglie particle in the presence of a potential field $\phi $.

If a photon is permitted to possess a frequency-independent
effective rest mass $m_{\rm eff}$, then the expression
(\ref{eq34}) for the case of photons in media is rewritten
\begin{equation}
n^{2}\left( \omega\right) =1-\frac{m_{\rm eff}^{2}c^{4}}{\hbar
^{2}\omega ^{2}}-2\frac{\phi }{\omega }+\left( \frac{\phi }{\omega
}\right) ^{2}.            \label{eq35}
\end{equation}
In what follows, an approach to frequency-independent effective
mass extracted from the optical refractive index squared
$n^{2}\left( \omega\right) $ is presented. It is assumed that
$n^{2}\left( \omega\right) $ can be rewritten as the following
series expansions
\begin{equation}
n^{2}\left( \omega \right) =\sum_{k}\frac{a_{-k}}{\omega ^{k}},
\end{equation}
then one can arrive at $n^{2}\left( \omega \right) \omega
^{2}=\sum_{k}\frac{a_{-k}}{\omega ^{k-2}}$, namely,
\begin{equation}
n^{2}\left( \omega \right) \omega ^{2}= ...+\frac{a_{-3}}{\omega
}+a_{-2}+a_{-1}\omega +a_{0}\omega ^{2}+a_{+1}\omega ^{3}+...\quad
. \label{eq37}
\end{equation}
Compared (\ref{eq37}) with (\ref{eq35}), it is apparent that the
frequency-independent effective rest mass squared $m_{\rm
eff}^{2}$ of the photon can be read off from $n^{2}\left( \omega
\right)$, namely, the constant term on the right-handed side on
(\ref{eq37}) is related only to $m_{\rm eff}^{2}$, {\it i.e.},
\begin{equation}
a_{-2}=-\frac{m_{\rm eff}^{2}c^{4}}{\hbar ^{2}}.
\end{equation}

Generally speaking, the optical refractive index $n^{2}\left(
\omega \right) $ is often of complicated form, particularly for
the metamaterials. So, we cannot extract frequency-independent
$m_{\rm eff}^{2}$ immediately from $n^{2}\left( \omega \right) $.
In the following, this problem is resolved by calculating the
limit value of $n^{2}\left( \omega \right) \omega ^{2}$, where
$\omega $ tends to $\infty $ or zero. We first consider the large
frequency behavior of $n^{2}\left( \omega \right) \omega ^{2}$. It
follows from (\ref{eq37}) that $n^{2}\left( \omega \right) \omega
^{2}$ at high frequencies ({\it i.e.}, $\omega \rightarrow \infty
$) is of the form
\begin{equation}
n^{2}\left( \omega \right) \omega ^{2}{\rightarrow
}a_{-2}+a_{-1}\omega +a_{0}\omega ^{2}+a_{+1}\omega
^{3}+a_{+2}\omega ^{4}+...\quad .
\end{equation}
This, therefore, means that $n^{2}\left( \omega \right) \omega
^{2}$ at high frequencies ($\omega \rightarrow \infty $) can be
split into the following two terms
\begin{equation}
n^{2}\left( \omega \right) \omega ^{2}{\rightarrow }\left( {\rm
constant \quad term}\right) +\left( {\rm divergent \quad
term}\right).
\end{equation}
So, we can obtain $m_{\rm eff}^{2}$ from the following formula
\begin{equation}
\lim_{\omega \rightarrow \infty }\left[ {n^{2}\left( \omega
\right) \omega ^{2}-\left( {\rm divergent \quad
term}\right)}\right]=-\frac{m_{\rm eff}^{2}c^{4}}{\hbar ^{2}}.
\label{eq311}
\end{equation}
Note that the constant term in $n^{2}\left( \omega \right) \omega
^{2}$ can also be extracted by considering the low-frequency
(zero-frequency) behavior of $n^{2}\left( \omega \right) $. It is
certain that in this case the obtained constant term is the same
as (\ref{eq311}). Apparently $n^{2}\left( \omega \right) \omega
^{2}$ at low-frequencies ({\it i.e.}, $\omega \rightarrow 0$)
behaves like
\begin{equation}
n^{2}\left( \omega \right) \omega ^{2}\rightarrow
...+\frac{a_{-5}}{\omega ^{3}}+\frac{a_{-4}}{\omega
^{2}}+\frac{a_{-3}}{\omega }+a_{-2}, \label{eq312}
\end{equation}
where the sum term ($...+\frac{a_{-5}}{\omega
^{3}}+\frac{a_{-4}}{\omega ^{2}}+\frac{a_{-3}}{\omega }$) is
divergent if $\omega$ approaches zero. Thus, subtraction of the
divergent term from $n^{2}\left( \omega \right) \omega ^{2}$
yields
\begin{equation}
\lim_{\omega \rightarrow 0}\left[ n^{2}\left( \omega \right)
\omega ^{2}-\left( {\rm divergent \quad term}\right) \right]
=-\frac{m_{\rm eff}^{2}c^{4}}{\hbar ^{2}}.       \label{eq313}
\end{equation}

In the next section, we will calculate the frequency-independent
effective rest mass of photons in the 2TDLM model by making use of
the method presented above.
\section{Frequency-independent effective rest mass of photons in the 2TDLM model}
According to the definition of the 2TDLM model, the
frequency-domain electric and magnetic susceptibilities are
expressed by Eq.(\ref{eqq4}) and the refractive index squared is
given in Eq.(\ref{eq3})

In order to obtain the frequency-independent $m_{\rm eff}^{2}$
from $n^{2}\left( \omega \right)$, according to the formulation in
the previous section, the high-frequency ($\omega \rightarrow
\infty $) behavior of $\chi ^{\rm e}\left( \omega \right) \omega
^{2}$ should first be taken into consideration and the result is
\begin{equation}
\chi ^{\rm e}\left( \omega \right) \omega ^{2}\rightarrow -i\omega
\omega _{\rm p}^{\rm e}\chi _{\beta }^{\rm e}+\omega ^{2}\chi
_{\gamma }^{\rm e}.                   \label{eq43}
\end{equation}
Note that $-i\omega \omega _{\rm p}^{\rm e}\chi _{\beta }^{\rm
e}+\omega ^{2}\chi _{\gamma }^{\rm e}$ is the divergent term of
$\chi ^{\rm e}\left( \omega \right) \omega ^{2}$ at large
frequencies. Subtracting the divergent term from $\chi ^{\rm
e}\left( \omega \right) \omega ^{2}$, we consider again the
high-frequency behavior ($\omega \rightarrow \infty $) and obtain
another divergent term $i\omega ^{3}\Gamma ^{\rm e}\chi _{\gamma
}^{\rm e}$, {\it i.e.},
\begin{equation}
\chi ^{\rm e}\left( \omega \right) \omega ^{2}-\left( -i\omega
\omega _{\rm p}^{\rm e}\chi _{\beta }^{\rm e}+\omega ^{2}\chi
_{\gamma }^{\rm e}\right) =\frac{\omega ^{2}\left( \omega _{\rm
p}^{\rm e}\right) ^{2}\chi _{\alpha }^{\rm e}-\omega ^{2}\omega
_{\rm p}^{\rm e}\Gamma ^{\rm e}\chi _{\beta }^{\rm e}+i\omega
\omega _{\rm p}^{\rm e}\left( \omega _{0}^{\rm e}\right) ^{2}\chi
_{\beta }^{\rm e}-i\omega ^{3}\Gamma ^{\rm e}\chi _{\gamma }^{\rm
e}-\omega ^{2}\left( \omega _{0}^{\rm e}\right) ^{2}\chi _{\gamma
}^{\rm e}}{-\omega ^{2}+i\omega \Gamma ^{\rm e}+\left( \omega
_{0}^{\rm e}\right) ^{2}}\rightarrow i\omega ^{3}\Gamma ^{\rm
e}\chi _{\gamma }^{\rm e}. \label{eq44}
\end{equation}
Thus, after subtracting all the divergent terms from $\chi ^{\rm
e}\left( \omega \right) \omega ^{2}$, we obtain
\begin{equation}
\lim_{\omega \rightarrow \infty }\left[ \chi ^{\rm e}\left( \omega
\right) \omega ^{2}-\left( -i\omega \omega _{\rm p}^{\rm e}\chi
_{\beta }^{\rm e}+\omega ^{2}\chi _{\gamma }^{\rm e}\right)
-i\omega ^{3}\Gamma ^{\rm e}\chi _{\gamma }^{\rm e}\right]
=-\left[ \left( \omega _{\rm p}^{\rm e}\right) ^{2}\chi _{\alpha
}^{\rm e}-\omega _{\rm p}^{\rm e}\Gamma ^{\rm e}\chi _{\beta
}^{\rm e}-\left( \omega _{0}^{\rm e}\right) ^{2}\chi _{\gamma
}^{\rm e}+\left( \Gamma ^{\rm e}\right) ^{2}\chi _{\gamma }^{\rm
e}\right]. \label{eq45}
\end{equation}
In the same fashion, we obtain
\begin{equation}
\lim_{\omega \rightarrow \infty }\left[ \chi ^{\rm m}\left( \omega
\right) \omega ^{2}-\left( -i\omega \omega _{\rm p}^{\rm m}\chi
_{\beta }^{\rm m}+\omega ^{2}\chi _{\gamma }^{\rm m}\right)
-i\omega ^{3}\Gamma ^{\rm m}\chi _{\gamma }^{\rm m}\right]
=-\left[ \left( \omega _{\rm p}^{\rm m}\right) ^{2}\chi _{\alpha
}^{\rm m}-\omega _{\rm p}^{\rm m}\Gamma ^{\rm m}\chi _{\beta
}^{\rm m}-\left( \omega _{0}^{\rm m}\right) ^{2}\chi _{\gamma
}^{\rm m}+\left( \Gamma ^{\rm m}\right) ^{2}\chi _{\gamma }^{\rm
m}\right] \label{eq46}
\end{equation}
for the magnetic susceptibility $\chi ^{\rm m}\left( \omega
\right) $. In what follows we continue to consider the term $\chi
^{\rm e}\left( \omega \right) \chi ^{\rm m}\left( \omega \right) $
on the right-handed side of (\ref{eq3}). In the similar manner,
the infinite-frequency limit of $\chi ^{\rm e}\left( \omega
\right) \chi ^{\rm m}\left( \omega \right) \omega ^{2}$ is given
as follows
\begin{eqnarray}
\lim_{\omega \rightarrow \infty }\chi ^{\rm e}\left( \omega
\right) \chi ^{\rm m}\left( \omega \right) \omega ^{2}&=&-\chi
_{\gamma }^{\rm m}\left[ \left( \omega _{\rm p}^{\rm e}\right)
^{2}\chi _{\alpha }^{\rm e}-\omega _{\rm p}^{\rm e}\Gamma ^{\rm
e}\chi _{\beta }^{\rm e}-\left( \omega _{0}^{\rm e}\right)
^{2}\chi _{\gamma }^{\rm e}+\left( \Gamma ^{\rm e}\right)
^{2}\chi _{\gamma }^{\rm e}\right]                               \nonumber \\
&-&\chi _{\gamma }^{\rm e}\left[ \left( \omega _{\rm p}^{\rm
m}\right) ^{2}\chi _{\alpha }^{\rm m}-\omega _{\rm p}^{\rm
m}\Gamma ^{\rm m}\chi _{\beta }^{\rm m}-\left( \omega _{0}^{\rm
m}\right) ^{2}\chi _{\gamma }^{\rm m}+\left( \Gamma ^{\rm
m}\right) ^{2}\chi _{\gamma }^{\rm m}\right].    \label{eq47}
\end{eqnarray}
Hence, insertion of (\ref{eq45})-(\ref{eq47}) into (\ref{eq311})
yields
\begin{eqnarray}
\frac{m_{\rm eff}^{2}c^{4}}{\hbar ^{2}}&=&\left( 1+\chi _{\gamma
}^{\rm m}\right) \left[ \left( \omega _{\rm p}^{\rm e}\right)
^{2}\chi _{\alpha }^{\rm e}-\omega _{\rm p}^{\rm e}\Gamma ^{\rm
e}\chi _{\beta }^{\rm e}-\left( \omega _{0}^{\rm e}\right)
^{2}\chi _{\gamma }^{\rm e}+\left( \Gamma ^{\rm e}\right) ^{2}\chi
_{\gamma
}^{\rm e}\right]                \nonumber \\
&+&\left( 1+\chi _{\gamma }^{\rm e}\right)\left[ \left(\omega
_{\rm p}^{\rm m}\right) ^{2}\chi _{\alpha }^{\rm m}-\omega _{\rm
p}^{\rm m}\Gamma ^{\rm m}\chi _{\beta }^{\rm m}-\left( \omega
_{0}^{\rm m}\right) ^{2}\chi _{\gamma }^{\rm m}+\left( \Gamma
^{\rm m}\right) ^{2}\chi _{\gamma }^{\rm m}\right]. \label{eq48}
\end{eqnarray}
Thus the $\omega$-independent effective rest mass squared $m_{\rm
eff}^{2}$ is extracted from $n^{2}\left( \omega \right)$.
Apparently, $m_{\rm eff}^{2}$ in Eq. (\ref{eq48}) does not depend
on $\omega$. The remainder on the right-handed side of Eq.
(\ref{eq35}) is $n^{2}\left( \omega \right) -1+\frac{m_{\rm
eff}^{2}c^{4}}{\hbar ^{2}\omega ^{2}}$, which is equal to $\left(
\frac{\phi }{\omega }\right) ^{2}-2\frac{\phi }{\omega }$, where
$\phi$ is defined to be $\frac{V}{\hbar }$. For simplicity, we set
\begin{equation}
\Delta \left( \omega \right) =n^{2}\left( \omega \right)
-1+\frac{m_{\rm eff}^{2}c^{4}}{\hbar ^{2}\omega ^{2}},
\label{eq4111}
\end{equation}
and then from the equation
\begin{equation}
\left( \frac{\phi }{\omega }\right) ^{2}-2\frac{\phi }{\omega
}=\Delta \left( \omega \right),
\end{equation}
one can readily obtain the expression for the hypothetical
potential field $\phi \left( \omega \right)$ is $\phi \left(
\omega \right) =\omega \left[ 1- \sqrt{1+\Delta \left( \omega
\right) }\right]$, or
\begin{equation}
V\left( \omega \right) =\hbar \omega \left[ 1- \sqrt{1+\Delta
\left( \omega \right) }\right].  \label{eq411}
\end{equation}

Many metamaterials which have complicated expressions for electric
and magnetic susceptibilities such as (\ref{eqq4}) are generally
rarely seen in nature and often arises from the artificial
manufactures. More recently, one of the artificial composite
metamaterials, the left-handed medium which has a frequency band
(GHz) where the electric permittivity ($\epsilon $) and the
magnetic permeability ($\mu $) are simultaneously negative, has
focused attention of many authors both experimentally and
theoretically\cite{Smith,Pendry3,Shelby,Ziolkowski2,Jianqi,Pendry2}.
In the left-handed medium, most phenomena as the Doppler effect,
Vavilov-Cherenkov radiation and even Snell's law are inverted. In
1964\footnote{Note that, in the literature, some
authors\cite{Kong,Garcia} mentioned the wrong year when Veselago
suggested the {\it left-handed media}. They claimed that Veselago
proposed or introduced the concept of {\it left-handed media} in
1968 or 1964. On the contrary, the true history is as follows:
Veselago's excellent paper was first published in Russian in July,
1967 [Usp. Fiz. Nauk {\bf 92}, 517-526 (1967)]. This original
paper was translated into English by W.H. Furry and published
again in 1968 in the journal of Sov. Phys. Usp.\cite{Veselago}.
Unfortunately, Furry stated erroneously in his English translation
that the original version of Veselago' work was first published in
1964.}, Veselago first considered many peculiar optical and
electromagnetic properties, phenomena and effects in this medium
and referred to such materials as left-handed
media\cite{Veselago}, since in this case the propagation vector
$\bf k$, electric field $\bf E$ and magnetic field $\bf H$ of
light wave propagating inside it form a left-handed system. It
follows from Maxwellian curl equations that such media having
negative simultaneously negative $\epsilon $ and $\mu $ exhibit a
negative index of refraction, {\it i.e.}, $n=-\sqrt{\epsilon \mu
}$. Since negative refractive index occurs only rarely, this
medium attracts attention of many physicists in various fields. In
experiments, the negative $\epsilon $ and $\mu $ can be
respectively realized by using a network (array) of thin (long)
metal wires\cite{Pendry2} and a periodic arrangement of split ring
resonators\cite{Pendry3}. A combination of the two structures
yields a left-handed medium. Compared the permittivity and
permeability (\ref{eqq}) of left-handed medium with the electric
and magnetic susceptibilities (\ref{eqq4}), one can arrive at
\begin{eqnarray}
\chi _{\alpha }^{\rm e}=1, \quad \omega _{\rm p}^{\rm e}=\omega
_{\rm p}, \quad \chi _{\beta }^{\rm e}=\chi _{\gamma }^{\rm e}=0,
\quad \omega _{0}^{\rm e}=0, \quad  \Gamma ^{\rm e}=-\gamma
,                                         \nonumber \\
 \chi _{\gamma }^{\rm m}=-F, \quad \chi _{\alpha }^{\rm m}=\chi _{\beta
}^{\rm m}=0, \quad \omega _{0}^{\rm m}=\omega _{0}, \quad \Gamma
^{\rm m}=-\Gamma.    \label{eq413}
\end{eqnarray}
So, it is readily verified with the help of (\ref{eq48}) that the
$\omega$-independent effective rest mass of photons inside
left-handed media is written as follows
\begin{equation}
\frac{m_{\rm eff}^{2}c^{4}}{\hbar ^{2}}=\left( 1-F\right) \omega
_{\rm p}^{2}+F\left(\omega _{0}^{2}-\Gamma ^{2}\right).
\label{eq414}
\end{equation}

It follows that Eq.(\ref{eq414}) is a restriction on the
electromagnetic parameters $F, \omega _{\rm p},  \omega _{0},
\Gamma $ and $F$ in the electric permittivity and magnetic
permeability (\ref{eqq}), since $m_{\rm eff}^{2} \geq 0$.
Insertion of the experimentally chosen values of the
electromagnetic parameters in the literature into Eq.
(\ref{eq414}) shows that this restriction condition is satisfied.
For instance, in Ruppin's work\cite{Ruppin}, $F, \omega _{\rm p},
\omega _{0}, \gamma, \Gamma $ and $F$ were chosen $F=0.56$,
$\omega _{\rm p}=10.0 {\rm GHz}$, $\omega _{0}=4.0 {\rm GHz}$,
$\gamma=0.03\omega _{\rm p}$, $ \Gamma=0.03\omega _{0}$, which
indicates that the inequality $m_{\rm eff}^{2} \geq 0$ is
satisfied.
\section{Some applications of frequency-independent effective rest mass of photons}
In this section we consider the applications of the concept of
frequency-independent effective rest mass of photons to some
optical and electromagnetic phenomena and effects.

(i) As was stated above, since the frequency-independent effective
rest mass of photons in electromagnetic media is related close to
the coupling parameters of the electric (magnetic) fields to the
electric (magnetic) dipole systems, and the damping and resonance
frequency of the electric (magnetic) dipole oscillators, it
contains the information on the interaction between light and
materials. It is shown that in certain frequency ranges this
effective rest mass governs the propagation behavior of
electromagnetic wave in media, and by using this concept one can
therefore consider the wave propagation somewhat conveniently. For
example, in the left-handed media whose permittivity and
permeability are expressed by (\ref{eqq}), the potential field
$V(\omega)$ approaches
$-\hbar\sqrt{F(\omega_{0}^{2}-\omega^{2}_{\rm p})}$, which is
constant, as the frequency tends to zero. This, therefore, means
that at low frequencies the light propagation is governed mainly
by the frequency-independent effective rest mass of photons.

Likewise, in the case of the potential field $V(\omega)$ being
vanishing where the frequency squared of electromagnetic wave is
\begin{equation}
\Omega^{2}=\frac{\omega^{2}_{0}\pm \sqrt{4\omega^{2}_{\rm
p}\omega_{0}^{2}-3\omega_{0}^{4}}}{2},       \label{51}
\end{equation}
the photon can be regarded as a ``free'' massive particle with the
rest mass being expressed in (\ref{eq414}). In the frequency
region near $\Omega$, sine the photon is a ``quasi-free''
particle, the light propagation behavior and properties is rather
simple compared with other frequency ranges in which $V(\omega)$
is nonvanishing or large. For the left-handed media, the relevant
frequency parameters $\omega_{0}$ and $\omega_{\rm p}$ of which is
about GHZ, the frequency $\Omega$ in Eq.(\ref{51}) that leads to
$V(\Omega)=0$ is just in the frequency band (GHZ) where the
permittivity and permeability are simultaneously negative. It
follows that when the effects and phenomena associated with
negative index of refraction occur in left-handed media, the wave
propagation ({\it e.g.}, scattering properties) and dispersion
properties of the light are not as complicated as those in other
frequency ranges, since in the negative-index frequency band, the
optical refractive index squared simulates nearly the plasma
behavior, particularly in the region near $\Omega$.

(ii) Historically, possibility of a light pulse with speed greater
than that ($c$) in a vacuum has been extensively investigated by
many authors\cite
{Ziolkowski,Akulshin,Bolda,Mitchell,Zhou,Nimtz1,Wang}. Ziolkowski
has studied the superluminal pulse propagation and consequent
superluminal information exchange in the 2TDLM media, and
demonstrated that they do not violate the principle of
causality\cite {Ziolkowski}. He showed that in the 2TDLM model,
both the phase and group velocities, {\it i.e.},
\begin{equation}
\lim_{\omega\rightarrow \infty}v(\omega)\sim
\frac{c}{|1+\chi_{\gamma}|}
\end{equation}
of light at high frequencies will exceed the speed ($c$) of light
in a vacuum, so long as $-2<\chi_{\gamma}<0$, where
$\chi_{\gamma}$ denotes the coupling coefficients of the second
time derivatives of electric (magnetic) fields to the local
electric (magnetic) dipole motions\cite {Ziolkowski}.

Although in the 2TDLM model the medium will exhibit a possibility
of superluminal speeds of wave propagation, the photon velocity
(rather than group velocity of light) is by no means larger than
$c$. This may be explained as follows: according to
Eq.(\ref{eq411}), the kinetic energy of a photon is ${\mathcal
E}=\hbar \omega\sqrt{1+\Delta(\omega)}$. If the light frequency
$\omega$ tends to infinity, then it follows from Eq.(\ref{eq4111})
that ${\mathcal E}\rightarrow n\hbar\omega$. Since the photon
momentum $p=\frac{n\hbar\omega}{c}$, the velocity $v$ of the
photon with a frequency-independent rest mass approaches $c$, {\it
i.e.}, the particle (photon) velocity $v=\frac{p}{{\mathcal
E}}\rightarrow c$. Thus we show that in the artificial composite
metamaterials such as 2TDLM media the photon velocity does not
exceed the speed of light in a vacuum, no matter whether the phase
(and group) velocity of light inside it is a superluminal speed or
not.

(iii) The frequency-independent rest mass of photons is useful in
discussing the plus and minus signs of phase shifts in light
scattering inside media. The phase shifts and its signs contain
the information on the scattering between wave and potentials.
According to Eq.(\ref{eq411}), if $\Delta(\omega)$ is positive
(negative), then the interaction energy $V(\omega)$ between light
and media is negative (positive). For this reason, it is possible
for us to determine the signs (plus or minus) of phase shifts in
light scattering process by calculating $\Delta(\omega)$ and the
frequency-independent rest mass $m_{\rm eff}$. Now with the
development in design and fabrication of left-handed materials,
the zero-index ($\epsilon\sim 0, \mu\sim 0$) materials receive
attention in materials science and applied
electromagnetism\cite{Hu}. For this kind of media with the optical
refractive index $n\sim 0$, from Eq.(\ref{eq4111}) it follows that
\begin{equation}
\Delta(\omega)\simeq \frac{m_{\rm eff}^{2}c^{4}}{\hbar ^{2}\omega
^{2}}-1,
\end{equation}
{\it i.e.}, $\Delta(\omega)$ is approximately equal to
$\frac{m_{\rm eff}^{2}c^{4}}{\hbar ^{2}\omega ^{2}}-1$ and we can
determine the sign by comparing the light frequency with the
photon effective rest mass $m_{\rm eff}$. This, therefore, means
that if the frequency $\omega$ is larger than $\frac{m_{\rm
eff}c^{2}}{\hbar}$, then $\Delta<0$ and consequently the phase
shift of light wave due to scattering acquires a minus sign; and
if the light frequency $\omega$ is less than $\frac{m_{\rm
eff}c^{2}}{\hbar}$, then $\Delta>0$ and consequently the phase
shift acquires a plus sign.
\section{Effects of self-induced charge currents in the torsion balance experiment}
In this section, we consider the potential effects of self-induced
charge currents arising in the torsion balance
experiments\cite{Luo,Lakes}. The Lagrangian density of
electrodynamics reads
\begin{equation}
{\mathcal
L}=-\frac{1}{4}F_{\mu\nu}F_{\mu\nu}-\frac{1}{2}\mu^{2}_{\gamma}A_{\mu}A_{\mu}+\mu_{0}J_{\mu}A_{\mu},
\label{eqn21}
\end{equation}
where
$\mu_{\gamma}^{2}=\left({\frac{m_{\gamma}c}{\hbar}}\right)^{2}$
with $m_{\gamma}$, $\hbar$ and $c$ being the photon rest mass,
Planck's constant and speed of light in a free vacuum,
respectively, and $\mu_{0}$ denotes the magnetic permeability in a
vacuum and the sum over the repeated indices is implied. It
follows from (\ref{eqn21}) that the canonical momentum density is
written
\begin{equation}
\pi_{\mu}=\frac{\partial{\mathcal L}}{\partial\dot{A}_{\mu}},
\quad               \vec{\pi}(x)=-{\bf E}(x).
\end{equation}
In the electron plasma, due to the conservation law of the
canonical momentum density, {\it i.e.}, $\frac{{\rm d}}{{\rm
d}t}(m_{\rm e}{\bf v}+e{\bf A})=0$, we have ${\bf
v}=-\frac{e}{m_{\rm e}}{\bf A}+{\bf C}$ with ${\bf C}$ being a
constant velocity. By using the formula ${\mathcal H}=-{\bf
E}\cdot{\dot{\bf A}}-{\mathcal L}$ and electric current density
${\bf J}=Ne{\bf v}=-\frac{N{e}^{2}}{m_{\rm e}}{\bf A}+Ne{\bf C}$,
one can arrive at (in SI)
\begin{equation}
{\mathcal H}=\frac{1}{2\mu_{0}}\left[{\frac{{\bf
E}^{2}}{c^{2}}+{\bf B}^{2}+\mu_{\gamma}^{2}{\bf
A}^{2}+2\frac{\mu_{0}N{e}^{2}}{m_{\rm e}}{\bf
A}^{2}+\frac{1}{\mu_{\gamma}^{2}c^{2}}\left(\nabla\cdot{\bf
E}-\frac{{\bf J}_{0}}{\epsilon_{0}}\right)^{2}}\right]+{\bf
J}_{0}{\bf A}_{0}-Ne{\bf C}\cdot{\bf A}, \label{eqn23}
\end{equation}
where ${\bf J}_{0}=\rho$ and ${\bf A}_{0}=\phi$ are respectively
the electric charge density and electric scalar potential, and
$\epsilon_{0}$ represents the electric permittivity in a vacuum
and $c=\frac{1}{\sqrt{\epsilon_{0}\mu_{0}}}$. Note that here
$\frac{1}{\mu_{\gamma}^{2}c^{2}}\left(\nabla\cdot{\bf
E}-\frac{{\bf J}_{0}}{\epsilon_{0}}\right)^{2}$ results from both
the mass term ${\frac{\mu_{\gamma}^{2}}{c^{2}}}{\bf A}_{0}^{2}$
and the Gauss's law $\nabla\cdot{\bf E}=-\mu_{\gamma}^{2}{\bf
A}_{0}+\frac{{\bf J}_{0}}{\epsilon_{0}}$.

It follows from (\ref{eqn23}) that the {\it total effective rest
mass squared} of electromagnetic fields in electron plasma is
given as follows
\begin{equation}
\mu_{\rm tot}^{2}=\mu_{\gamma}^{2}+2\frac{\mu_{0}N{e}^{2}}{m_{\rm
e}}.                                          \label{eqn24}
\end{equation}
It is worthwhile to point out that according to the
Amp\`{e}re-Maxwell-Proca equation $\nabla\times{\bf
B}=\mu_{0}\epsilon_{0}\frac{\partial}{\partial t }{\bf
E}+\mu_{0}{\bf J}-\mu^{2}_{\gamma}{\bf A}$ and the consequent
$\nabla\times{\bf B}=\mu_{0}\epsilon_{0}\frac{\partial}{\partial t
}{\bf E}-[\mu^{2}_{\gamma}+\frac{\mu_{0}N{e}^{2}}{m_{\rm e}}]{\bf
A}+\mu_{0}Ne{\bf C}$, the $\mu_{\rm tot}^{2}$ should be $\mu_{\rm
tot}^{2}=\mu_{\gamma}^{2}+\frac{\mu_{0}N{e}^{2}}{m_{\rm e}}$
rather than that in (\ref{eqn24}). This minor difference between
these two $\mu_{\rm tot}^{2}$ results from the derivative
procedure applied to the Lagrangian density by using the
Euler-Lagrange equation. Although it seems from the
Amp\`{e}re-Maxwell-Proca equation that $\mu_{\rm
tot}^{2}=\mu_{\gamma}^{2}+\frac{\mu_{0}N{e}^{2}}{m_{\rm e}}$, the
factor-$2$ in (\ref{eqn24}) cannot be ignored in calculating the
torque acting on the torsion balance due to the photon effective
rest mass, which will be confirmed in what follows. In the
classical electromagnetics, the following familiar formulae are
given
\begin{equation}
\nabla\times{\bf B}=\mu_{0}{\bf J},  \quad   m_{d}=\pi r^{2}I,
\quad          I={\bf J}\cdot {\bf S},  \quad \vec{\tau}={\bf
m}_{d}\times {\bf B} \label{eqn25}
\end{equation}
and
\begin{equation}
\nabla\times{\bf A}={\bf B},  \quad  a_{d}=\pi
r^{2}\frac{\Phi}{\mu_{0}},  \quad   \Phi={\bf B}\cdot{\bf S},
\label{eqn26}
\end{equation}
where $\nabla\times{\bf B}=\mu_{0}{\bf J}$ holds when the electric
field strength ${\bf E}$, electric current density ${\bf J}$ are
vanishing in the Amp\`{e}re-Maxwell equation. In Eq.(\ref{eqn25}),
$m_{d}=\pi r^{2}I$ means that a current loop of radius $r$
carrying current $I$ gives rise to a magnetic dipole moment
$m_{d}$; ${\bf S}$ in $I={\bf J}\cdot {\bf S}$ denotes the area
vector through which the electric current density vector ${\bf J}$
penetrates; $\vec{\tau}={\bf m}_{d}\times {\bf B}$ means that an
electric-current loop with a magnetic dipole moment ${\bf m}_{d}$,
immersed in a magnetic field ${\bf B}$, experiences a torque
$\vec{\tau}={\bf m}_{d}\times {\bf B}$. Eqs.(\ref{eqn26}) means in
the torsion balance experiment, a toroid coil contains a loop of
magnetic flux ${\Phi}$ which acts as a dipole source $a_{d}$ of
magnetic vector potential via $\nabla\times{\bf A}={\bf B}$ with
the magnetic field ${\bf B}$ within the toroid as the source term.

It follows from (\ref{eqn23}) that in the Hamiltonian density
${\mathcal H}$ of electromagnetic system, there are
$\frac{1}{2}{\bf B}^{2}$ and
$\frac{1}{2}\left(\mu_{\gamma}^{2}+2\frac{\mu_{0}N{e}^{2}}{m_{\rm
e}}\right){\bf A}^{2}$ ({\it i.e.}, $\frac{1}{2}\left(\mu_{\rm
tot}{\bf A}\right)^{2}$), which provide us with some useful
insights into calculating the torque produced by the interaction
of magnetic dipole vector potential moment ${\bf a}_{d}$ with the
ambient cosmic magnetic vector potential acting upon the toroid in
the torsion balance experiment. For convenience, the Equations
(\ref{eqn26}) may be rewritten as follows
\begin{equation}
\nabla\times(\mu_{\rm tot}{\bf A})=\mu_{0}\left(\frac{\mu_{\rm
tot}{\bf B}}{\mu_{0}}\right),     \quad        \mu_{\rm
tot}a_{d}=\pi r^{2}\left(\frac{\mu_{\rm tot}
\Phi}{\mu_{0}}\right),                       \quad \frac{\mu_{\rm
tot}\Phi}{\mu_{0}}=\left(\frac{\mu_{\rm tot}{\bf
B}}{\mu_{0}}\right)\cdot{\bf S}.                    \label{eqn27}
\end{equation}

In order to calculate the torque acting on the torsion balance, we
compare the equations and expressions in (\ref{eqn27}) with those
in (\ref{eqn25}) as follows:
$$
\vbox{\tabskip=0pt \offinterlineskip \halign to \hsize {\strut#&
\vrule#\tabskip=1em plus2em &\hfil#\hfil& \vrule# &\hfil#\hfil&
\vrule#&\hfil#\hfil & \vrule#\tabskip=0pt\cr \noalign{\hrule}

&& \omit\hidewidth  \hidewidth & & \omit\hidewidth the familiar
case \hidewidth && \omit\hidewidth the case due to photon mass
\hidewidth & \cr \noalign{\hrule} && in ${\mathcal H}$ &&
$\frac{1}{2}{\bf B}^{2}$ &&$\frac{1}{2}\left(\mu_{\rm tot}{\bf
A}\right)^{2}$&  \cr \noalign{\hrule} && Equations &&
$\nabla\times{\bf B}=\mu_{0}{\bf J}$ && $\nabla\times(\mu_{\rm
tot}{\bf A})=\mu_{0}\left(\frac{\mu_{\rm tot}{\bf
B}}{\mu_{0}}\right)$& \cr \noalign{\hrule} && dipole moment &&
$m_{d}=\pi r^{2}I$ && $ \mu_{\rm tot}a_{d}=\pi
r^{2}\left(\frac{\mu_{\rm tot} \Phi}{\mu_{0}}\right)$& \cr
\noalign{\hrule} && flux && $I={\bf J}\cdot {\bf S}$ && $\quad
\frac{\mu_{\rm tot}\Phi}{\mu_{0}}=\left(\frac{\mu_{\rm tot}{\bf
B}}{\mu_{0}}\right)\cdot{\bf S}$& \cr \noalign{\hrule} && torque
&& $\vec{\tau}={\bf m}_{d}\times {\bf B}$ && $\vec{\tau}=?$ &  \cr
\noalign{\hrule} \noalign{\smallskip}
 \multispan7*{Table {\it 1}}
\hfil\cr}}
$$

Thus it follows from the analogies in the above table that,
$?=(\mu_{\rm tot}{\bf a}_{d})\times(\mu_{\rm tot}{\bf A}) $, {\it
i.e.}, the torque $\vec{\tau}=\mu_{\rm tot}^{2}{\bf
a}_{d}\times{\bf A}$. Here the torque $\vec{\tau}$ arising from
the interaction between the dipole field of magnetic potentials
(resulting from the electric current $I$ in the toroid windings)
with the ambient vector potential ${\bf A}$ acts on the torsion
balance. It should be noted again that here $\mu_{\rm
tot}^{2}=\mu_{\gamma}^{2}+2\frac{\mu_{0}N{e}^{2}}{m_{\rm e}}$
rather than $\mu_{\rm
tot}^{2}=\mu_{\gamma}^{2}+\frac{\mu_{0}N{e}^{2}}{m_{\rm e}}$.
\\

Note that there exists a term $\mu_{0}Ne{\bf C}$ in the
Amp\`{e}re-Maxwell-Proca equation $\nabla\times{\bf
B}=\mu_{0}\epsilon_{0}\frac{\partial}{\partial t }{\bf
E}-[\mu^{2}_{\gamma}+\frac{\mu_{0}N{e}^{2}}{m_{\rm e}}]{\bf
A}+\mu_{0}Ne{\bf C}$. Does this term has influences on the torsion
balance? In what follows we will discuss this problem.

In the similar fashion, we consider the familiar Hamiltonian
density $-{\bf M}\cdot{\bf B}$, which describes the interaction of
the magnetic moments with the magnetic field ${\bf B}$. According
to the Amp\`{e}re-Maxwell-Proca equation with
$\mu_{0}\epsilon_{0}\frac{\partial}{\partial t }{\bf E}$ and
$\frac{\mu_{0}N{e}^{2}}{m_{\rm e}}{\bf A}$ being ignored, we have
$\nabla\times{\bf B}=\mu_{0}Ne{\bf C}$, which can be rewritten as
$\nabla\times{\bf B}=\left(\mu_{0}\mu_{\rm
tot}\right)\left(\frac{Ne{\bf C}}{\mu_{\rm tot}}\right)$. In the
meanwhile,  ${\mathcal H}=-Ne{\bf C}\cdot{\bf A}$ can also be
rewritten as ${\mathcal H}=-\left(\frac{Ne{\bf C}}{\mu_{\rm
tot}}\right)\cdot\left(\mu_{\rm tot}{\bf A}\right)$.

It is known that in the sufficiently large space filled with
homogeneous magnetic moments with volume density (magnetization)
being ${\bf M}$, the interior magnetic field strength ${\bf B}$
produced by homogeneously distributed magnetic moments is just
$\mu_{0}{\bf M}$\footnote{This may be considered in two ways:
({\rm i}) if the direction of ${\bf M}$ is assumed to be parallel
to the third component of Cartesian coordinate system, then the
potentially non-vanishing components of ${\bf B}$ and ${\bf H}$
are also the third ones. It follows from the Amp\`{e}re's law
$\nabla\times{\bf H}={\bf J}_{\rm free}$ with ${\bf J}_{\rm
free}=0$ that $H_{3}$ is constant, and $H_{3}$ can be viewed as
zero. So, according to the expression ${\bf B}=\mu_{0}({\bf
H+M})$, we obtain ${\bf B}=\mu_{0}{\bf M}$; ({\rm ii}) In
accordance with Amp\`{e}re's circuital law, the magnetic induction
$B$ inside a long solenoid with $N$ circular loops carrying a
current $I$ in the region remote from the ends is axial, uniform
and equal to $\mu_{0}$ times the number of Amp\`{e}re-turns per
meter $nI$, {\it i.e.}, $B=\mu_{0}nI$, where $n=\frac{N}{l}$ with
$l$ being the solenoid length. Let $S$ denote the cross-section
area of this long solenoid. The volume density of magnetic moments
$IS$ resulting from the carried current $I$ is
$M=\frac{(IS)nl}{V}$ with the solenoid volume $V=Sl$. This,
therefore, means that the volume density of magnetic moments just
equals the number of Amp\`{e}re-turns per meter $nI$, {\it i.e.},
$M=nI$. So, we have $B=\mu_{0}M$.}, {\it i.e.}, $\nabla\times{\bf
A}=\mu_{0}{\bf M}$ or $\nabla\times\left({\mu_{\rm tot}}{\bf
A}\right)=\left(\mu_{0}{\mu_{\rm tot}}\right){\bf M}$.

Thus the comparisons between ``{\it the familiar case}'' and {\it
the case due to the constant} ${\bf C }$ are illustrated in the
following table,
$$
\vbox{\tabskip=0pt \offinterlineskip \halign to \hsize {\strut#&
\vrule#\tabskip=1em plus2em &\hfil #\hfil & \vrule#&\hfil#\hfil&
\vrule#&\hfil#\hfil & \vrule#\tabskip=0pt\cr \noalign{\hrule}

&& \omit\hidewidth  \hidewidth & & \omit\hidewidth the familiar
case \hidewidth && \omit\hidewidth the case due to constant ${\bf
C }$ \hidewidth & \cr \noalign{\hrule} && in ${\mathcal H}$ &&
$\frac{1}{2}{\bf B}^{2}$ &&$\frac{1}{2}\left(\mu_{\rm tot}{\bf
A}\right)^{2}$&  \cr \noalign{\hrule} && interaction Hamiltonian
density && ${\mathcal H}=-{\bf M}\cdot{\bf B}$ && ${\mathcal
H}=-\left(\frac{Ne{\bf C}}{\mu_{\rm
tot}}\right)\cdot\left(\mu_{\rm tot}{\bf A}\right)$& \cr
\noalign{\hrule} && Equations &&$\nabla\times\left({\mu_{\rm
tot}}{\bf A}\right)=\left(\mu_{0}{\mu_{\rm tot}}\right){\bf M}$ &&
$\nabla\times{\bf B}=\left(\mu_{0}\mu_{\rm
tot}\right)\left(\frac{Ne{\bf C}}{\mu_{\rm tot}}\right)$&  \cr
\noalign{\hrule} && torque && $\vec{\tau}={\bf M}\times{\bf B}$&&
$\vec{\tau}=?$& \cr \noalign{\hrule}  \noalign{\smallskip}
 \multispan7*{Table {\it 2}}
\hfil\cr}}
$$

It follows that $?=\left(\frac{Ne{\bf C}}{\mu_{\rm
tot}}\right)\times\left(\mu_{\rm tot}{\bf A}\right)$, {\it i.e.},
$\vec{\tau}=Ne{\bf C}\times{\bf A}$, which dose not act on the
toroid in the torsion balance experiment.
\\

Since we have discussed the related preliminary preparation for
the effects of self-induced charge currents, we think the order of
magnitude of its effects in experiments ({\it e.g.}, the extra
torque due to the photon effective mass acting on the toroid in
the torsion balance experiment) deserves evaluation. It is shown
by our evaluation that the muons and alpha-particles in secondary
cosmic rays will contribute an effective rest mass about
$10^{-54}$ Kg to the photon, which is compared to the newly
obtained upper limit on photon rest mass in Luo's {\it rotating
torsion balance} experiment.
\section{Discussion and remarks: self-induced charge currents in dilute plasma}
Here we discuss the effective rest mass of photon resulting from
the self-induced charge currents in the torsion balance
experiments\cite{Luo,Lakes}. The self-induced charge current
arises mainly from the two sources around us: ({\rm i}) muon
($\mu$) component and alpha-particles in cosmic rays; ({\rm ii})
decay daughter (alpha-particles) of radioactive radon gas
($^{222}_{86}{\rm Rn}$, $^{220}_{86}{\rm Rn}$) in room
environment.

It is known that at the sea level, the current density of muon
component in secondary cosmic rays is about $1\times 10^{-2}{\rm
cm}^{-2}\cdot{\rm s}^{-1}$\cite{Mei}. Assuming that the muon
velocity approaches speed of light, the volume density of muon can
be derived and the result is $N_{\mu}=0.3\times 10^{-6}$ ${\rm
m}^{-3}$. So, according to Eq.(\ref{eqn24}) ({\it i.e.}, $m_{\rm
eff}=\frac{\hbar}{c}\sqrt{\frac{2N_{\mu}e^{2}}{\epsilon_{0}m_{\mu}c^{2}}}$),
electromagnetic wave with wavelength\footnote{Since the mean
distance of two muons in secondary cosmic rays at sea level is
about 100 m, Lorentz's mean-field formulation (Lorentz's
dispersion theory, Lorentz's electron theory, 1895) is not
applicable to the electromagnetic wave with wavelength $\lambda
\ll 100$ m.  So, the above effective mass formula is no longer
valid for this case ({\it i.e.}, $\lambda \ll 100$ m). In fact,
the electromagnetic wave with wavelength $\lambda \ll 100$ m dose
not acquire this effective rest mass. But the ambient cosmic
magnetic vector potentials (interstellar magnetic fields) with low
(or zero) frequencies will truly acquire this effective rest
mass.} $\lambda \gg 100$ m at the sea level acquires an effective
rest mass about $0.3\times 10^{-53}$ Kg.

It is believed that the air molecules in the vacuum chamber of
torsion balance experiment cannot be ionized by the
alpha-particles of cosmic rays, because the mean free path of
alpha-particle moving at about $10^{7}{\rm m}\cdot{\rm s}^{-1}$ in
the dilute air with the pressure being only $10^{-2}$ Pa\cite{Luo}
is too large (more than $10^{6}$ m)\footnote{According to some
handbooks, the free paths of alpha-particles in air (with pressure
$10^{5}$ Pa) are 2.5, 3.5, 4.6, 5.9, 7.4, 8.9, 10.6 cm,
corresponding respectively to the energy 4.0, 5.0, 6.0, 7.0, 8.0,
9.0, 10.0 Mev. So, in the vacuum chamber with pressure $10^{-2}$
Pa in Luo's experiment, the free path of alpha-particles (with
energy much more than 10 Mev, even perhaps more than 100 Mev) in
secondary cosmic rays is more than $10^{7}$ times several cm,
which means that the air molecules in the low-pressure vacuum
chamber cannot be easily ionized by the alpha-particles of cosmic
rays.}. So, the air medium has nothing to contribute to the
effective rest mass of photons.

Radon ($_{86}{\rm Rn}$) possesses the $\alpha$-radioactivity.
Generally speaking, the Rn density in room environment is
$10^{-8}{\rm m}^{-3}$ or so, which may be the same order of
magnitude as particle density in secondary cosmic
rays\footnote{The effective doses of cosmic rays and Rn humans
suffer are respectively $0.4$ mSv (typical range: 0.3-1.0 mSv) and
$1.2$ mSv (typical range: 0.2-10 mSv).}, and the decay daughter
(alpha-particles) will therefore give rise to a photon effective
mass of about $10^{-54}$ Kg.
\\

Thus the effective rest mass induced by muons and alpha-particles
in secondary cosmic rays and radon gas can be compared to the new
upper limit ($1.2\times 10^{-54}$ Kg) obtained by Luo {\it et al.}
in their recent rotating torsion balance experiment\cite{Luo}. So,
as far as the contribution of ions of cosmic rays to the photon
rest mass is concerned, this upper limit ($1.2\times 10^{-54}$ Kg)
becomes just a critical value. We think Luo's experimentally
obtained upper limit is of much interest and therefore deserves
further experimental investigation to get a more precise upper
limit on photon mass.

\section{Conclusion}
In this paper,
\\

{\bf i}) it is verified that the expression $\frac{m_{\rm
eff}^{2}c^{4}}{\hbar ^{2}}=(1-n^{2})\omega ^{2}$ can be applied to
calculating the effective rest mass of photons in the artificial
composite electromagnetic metamaterials with complicated
permittivity and permeability.
\\

{\bf ii}) we suggest an approach to obtaining the
frequency-independent effective rest mass of photons in certain
artificial composite metamaterials such as 2TDLM media which
possess the complicated electric permittivity and magnetic
permeability.

{\bf iii}) it is shown that the extra torque
$2\frac{\mu_{0}Ne^{2}}{m}{\bf a}_{d}\times {\bf A}$ by the
self-induced charge currents will also acts upon the toroid in the
torsion balance experiment whose apparatus is immersed in the
dilute plasma such as the cosmic rays and radon gas. Here the
photon effective rest mass squared is $\mu_{\rm
eff}^{2}=2\frac{\mu_{0}N{e}^{2}}{m_{\rm e}}$ rather than $\mu_{\rm
eff}^{2}=\frac{\mu_{0}N{e}^{2}}{m_{\rm e}}$.

The following difference between these two $\mu_{\rm eff}^{2}$
should be pointed out: the theoretical mechanism of torsion
balance experiment lies in the Lagrangian or Hamiltonian
(\ref{eqn23}) of interacting electromagnetic system, so, the
effective rest mass squared of electromagnetic fields is $\mu_{\rm
eff}^{2}=2\frac{\mu_{0}N{e}^{2}}{m_{\rm e}}$. However, for those
experimental schemes\cite{Feinberg,Fischbach,Chernikov} of photon
rest mass stemming from the Amp\`{e}re-Maxwell-Proca equation
$\nabla\times{\bf B}=\mu_{0}\epsilon_{0}\frac{\partial}{\partial t
}{\bf E}-[\mu^{2}_{\gamma}+\frac{\mu_{0}N{e}^{2}}{m_{\rm e}}]{\bf
A}+\mu_{0}Ne{\bf C}$, the effective rest mass squared of
electromagnetic fields is $\mu_{\rm
eff}^{2}=\frac{\mu_{0}N{e}^{2}}{m_{\rm e}}$ rather than $\mu_{\rm
eff}^{2}=2\frac{\mu_{0}N{e}^{2}}{m_{\rm e}}$. The so-called
experimental realizations stemming from the
Amp\`{e}re-Maxwell-Proca equation are as follows: pulsar test of a
variation of the speed of light with frequency\cite{Feinberg};
geomagnetic limit on photon mass based on the analysis of
satellite measurements of the Earth's field\cite{Fischbach};
experimental test of Amp\`{e}re's law at low
temperature\cite{Chernikov}, {\it etc.}. In all these experimental
schemes, the effective mass squared is $\mu_{\rm
eff}^{2}=\frac{\mu_{0}N{e}^{2}}{m_{\rm e}}$.
\\

{\bf iv}) the self-induced charge currents in dilute plasma
presented above ({\it i.e.}, muons and alpha-particles in
secondary cosmic rays and radon gas) will give arise to an
effective rest mass of about $10^{-54}$ Kg, which can be compared
to the new upper limit on photon rest mass obtained in Luo's
rotating torsion balance experiment.
\\ \\

\textbf{Acknowledgements}  This project is supported in part by
the National Natural Science Foundation of China under the project
No. $90101024$.

\end{document}